# Study of Supersolidity and Shear Modulus Anomaly of $^4$He in a Triple Compound Oscillator

**Xiao Mi, Erich Mueller, and John D. Reppy**

Laboratory of Atomic and Solid State Physics, Cornell University, Ithaca, New York, 14853-2501, USA

Email: jdr13@cornell.edu

**Abstract**: The recently discovered shear modulus anomaly in solid $^4$He bears a strong similarity to the phenomenon of supersolidity in solid $^4$He and can lead to the period shift and dissipative signals in torsional oscillator experiments that are nearly identical to the classic NCRI signals observed by Kim and Chan. In the experiments described here, we attempt to isolate the effects of these two phenomena on the resonance periods of torsion oscillators. We have constructed a triple compound oscillator with distinct normal modes. We are able to demonstrate that, for this oscillator, the period shifts observed below 200 mK have their primary origin in the temperature dependence of the shear modulus of the solid $^4$He sample rather than the formation of a supersolid state.

## 1. Introduction:

In a series of landmark torsional oscillator (TO) experiments Kim and Chan (KC) [1,2], discovered the first experimental evidence for the existence of a supersolid state of matter in which solid $^4$He behaves simultaneously as superfluid and crystalline solid. In a typical supersolid TO experiment, the resonance period of a torsional oscillator containing a sample of solid $^4$He is observed to decrease below an onset temperature of about 200mK and is accompanied by a peak in the TO dissipation where the period is changing most rapidly as a function of temperature. These phenomena have been interpreted by KC as evidence for a Non-Classical Rotational Inertia (NCRI) [3], a superfluid-like decoupling of a fraction of solid $^4$He from the motion of the oscillator.

This interpretation is, however, challenged by a series of experiments done by Day and Beamish (DB) [4] which show an anomalous increase in the shear modulus, $\mu$, of solid $^4$He at low temperatures. The temperature dependence of $\mu$ bears strong similarity to the NCRI signal observed by KC, thus suggesting an alternative explanation for the decrease in torsional oscillator periods. A recent plastic deformation experiment [5], showed that the high temperature values of the torsional oscillator period is increased by increasing the density of dislocations in solid $^4$He, while the low temperature period values remained fixed. This result contradicts the supersolid interpretation, which assumes that the sample is rigidly locked to the oscillator motion above the NCRI onset temperature, making the oscillator period insensitive to the level of disorder.

The work reported here isolates the effects of shear stiffening and supersolidity on torsional oscillations through the study of a triple compound oscillator which has the advantage of allowing us to probe the NCRI signal of the same sample at several different frequencies. The approach is similar to that employed in a series of compound oscillator experiments by the Rutgers group of Kojima [6] and more recently by the group at Royal-Holloway [7]. On the theoretical side, a discussion of the response of double oscillators containing solid $^4$He samples has been given in terms of a non-supersolid glassy response of the solid by the Los Alamos group of Graf et al [8].

## 2. Triple Oscillator Structure and Data:

To further investigate the period signals seen in the plastic flow deformation experiments [5], we have mounted a cell of similar design on a dummy oscillator structure to form a triple oscillator, as shown in Figure 1. In the equations of motion for the triple oscillator, the variables $\theta_1$, $\theta_2$, $\theta_3$ will correspond to the angular displacements of the internal torsion bob with moment of inertia $I_1 = 2.58$ gcm$^2$, the sample cell with moment of inertia $I_2 = 97$ gcm$^2$, and the dummy oscillator with moment of inertia $I_3 = 50$ gcm$^2$. The oscillator structure is mounted on a large mass with a moment of inertia on the order of $4 \times 10^3$ gcm$^2$, which is in turn thermally anchored to the mixing chamber of a dilution refrigerator. The oscillator's internal torsion bob is supported by an aluminium rod with torsion constant $k_0$. When the annulus is filled with solid $^4$He, the internal torsion bob is coupled to the cell by an additional contribution due to the shear modulus of the solid sample. The effective torsion constant is then, $k_1 = k_0 + k_\mu$, where $k_\mu$ is the contribution arising from the shear modulus of the solid sample contained in the region between the walls of the cell and the internal torsion bob. The quantity $k_\mu$ is linearly proportional to the temperature dependent shear modulus. It can be calculated from the shear modulus values [4] and the dimensions of the sample consisting of an annular region of height $h = 0.975$ cm, width $\Delta R = 0.051$ cm, mean radius $R = 0.707$ cm and two disk shaped regions at the top and bottom of the inner torsion bob with heights $\Delta D = 0.051$ cm, according to the formula below:

$$k_\mu = \mu \pi R^3 \left( \frac{2h}{\Delta R} + \frac{R}{\Delta D} \right) = G\mu,$$ with the geometric constant $G = 57.8$ cm$^3$ for the parameters of our geometry. Since $\Delta R \ll R$, we have approximated the curved geometry of the narrow gap annulus as a parallel plate configuration in deriving the expression for $k_\mu$.

The triple oscillator structure will have three resonant frequencies, which we designate as $f_-$, $f_+$, and $f_1$. The highest mode frequency, $f_1$, is strongly affected by the presence or absence of solid in the sample region. At 500 mK, $f_- = 514.3$ Hz, $f_+ = 1313.5$ Hz and $f_1$ ranges from about 4400 Hz when the sample volume is filled with superfluid to over 9500 Hz for a solid $^4$He sample. The magnitude of $f_1$ for the unfrozen cell also allows for an estimate of $k_0 = 1.97 \times 10^9$ dynescm/rad, and the magnitude of $f_1$ allows for an estimate of $k_\mu = 7.22 \times 10^9$ dynescm/rad, corresponding to a value for shear modulus $\mu = 1.25 \times 10^8$ dynes/cm at 500 mK.

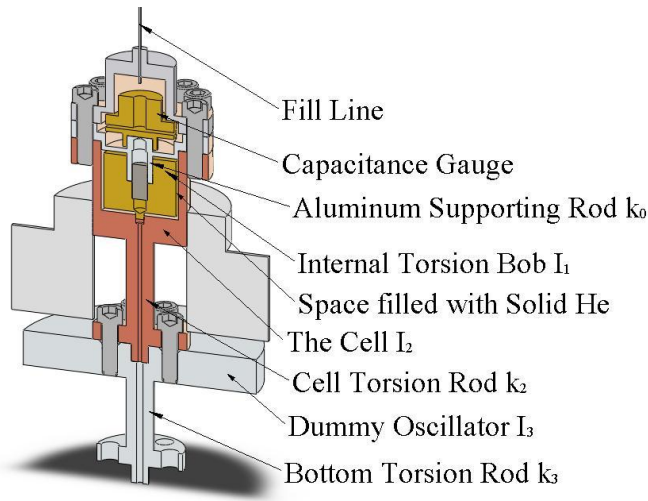

**Figure 1.** The Triple Compound Oscillator

In Figure 2 we show the temperature dependence of the periods and dissipation, $Q^{-1}$, for the $P_-$ and $P_+$ modes with the cell containing solid $^4$He at a pressure of about 38 bar. Data for the empty cell with period values shifted by a constant are shown for comparison. The temperature dependence of the periods for these modes and their accompanying dissipation peaks are of the classic NCRI form first seen by Kim and Chan. The sensitivity of the mode periods to mass loading has been determined by measuring the period changes, $\delta P$, in response to the addition of a small moment of inertia, $\Delta I$, to the body of the cell. The experimentally determined mass-loading sensitivities are $(\delta P_+/\Delta I) = 2.9 \times 10^{-6}$ sec/(gcm$^2$), $(\delta P_-/\Delta I) = 12.4 \times 10^{-6}$ sec/(gcm$^2$). The ratio of these two mass-loading sensitivities is $(\delta P_+/\delta P_-)_{mass-loading} = 0.234$.

Based on the dimensions of the sample volume, we estimate the moment of inertia of the solid to be $I_{He} = 2.62 \times 10^{-2}$ gcm$^2$ for a solid density of 0.2 gcm$^{-3}$. Thus, one might expect an increase in the periods upon freezing to be 76.1 ns and 325.4 ns for the $P_+$ and $P_-$ modes respectively. In the usual supersolid treatment these values are used as normalization factors in computing the supersolid fraction (NCRIF) from the period shift data such as shown in Figure 2.

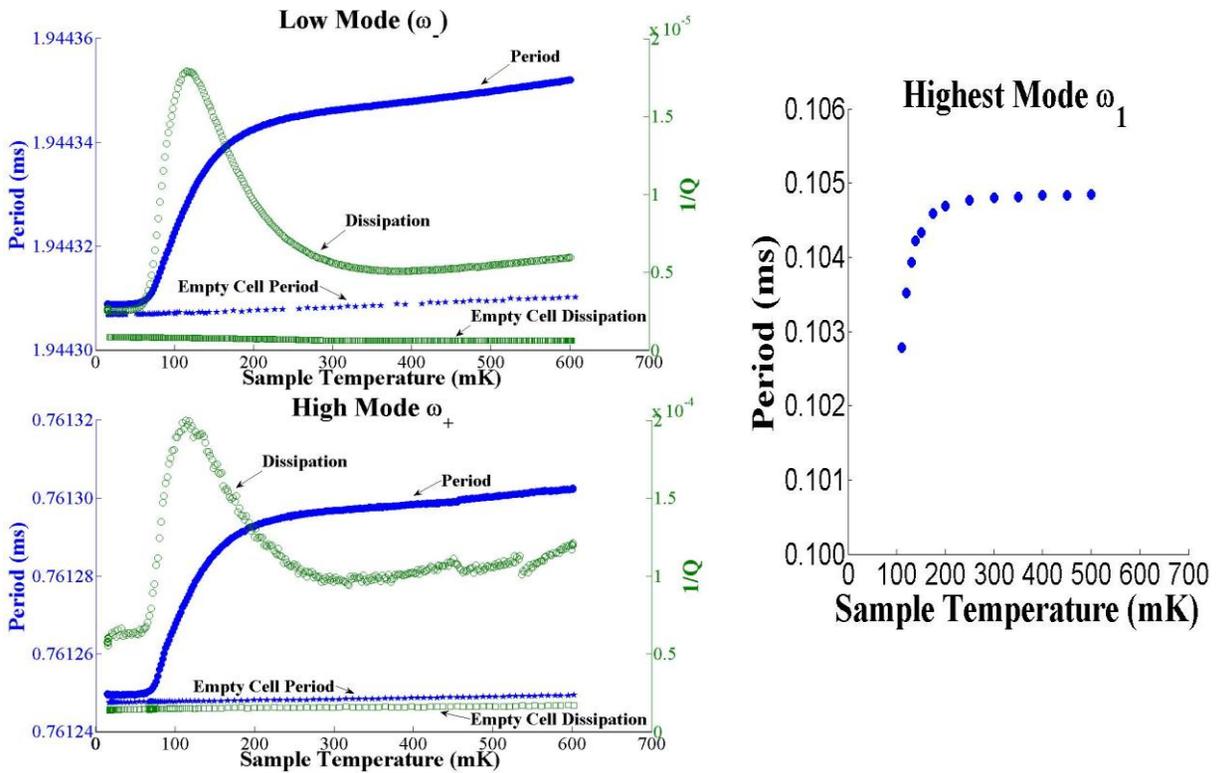

**Figure 2.** Periods and Dissipations of Triple Oscillator at the Three Normal Modes

## 3. Results

We can estimate the size of possible supersolid signals for these modes by extending the linear high temperature region to 20 mK and taking the difference between that extrapolated "background" and the actual period data. The period shifts obtained are 40 ns and 30 ns for the $P_+$ and $P_-$ modes respectively. Following the conventional supersolid analysis, these period shifts would correspond to NCRIF's of $52.6 \times 10^{-2}$ and $9.2 \times 10^{-2}$ for the + and – modes respectively. These results clearly violate the expectation of frequency independence for the NCRIF required in the classic supersolid scenario. We must, therefore, seek some other explanation for the observed dependence of the TO period on temperature.

In Figure 2 we have also included period data for the highest mode. Unfortunately we were only able to track this mode down to only 100 mK because of a rapidly decreasing $Q$. Over the temperature range of observation, however, the period of the highest mode shows a marked decrease, as expected for an increasing value of the shear modulus [4].

In Figure 3, we show another test of the data. Here we plot, starting at the lowest temperature, the increase in the period of the $P_+$ mode against the corresponding increase in the period of the $P_-$ mode. Over a temperature range extending from 20 mK to 600 mK the data obey, to good approximation, a linear relation, as would be expected if a single mechanism were in operation for both modes. In the case where the period shifts are due to temperature dependent mass-loading, as in the supersolid scenario, the data would follow a linear relation indicated by the lower dashed line. A second solid line, which is much closer to the experimental data, is based on an analysis, given below, of the triple oscillator, taking into account the variation of the shear modulus of the solid $^4$He. As will become evident, the temperature dependence of the period for these modes arises almost entirely from the temperature dependence of the shear modulus of the solid and a true supersolid signal, if present, must be relatively small.

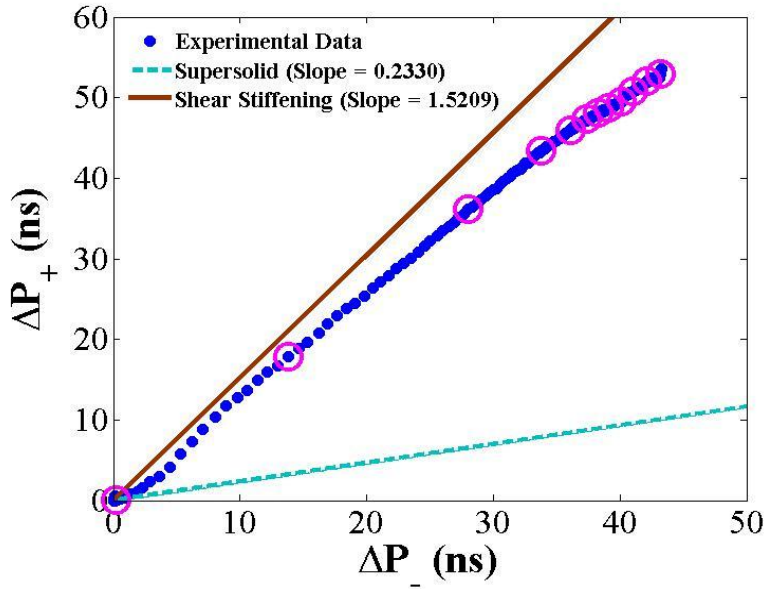

**Figure 3.** Plot of period shifts (with values at 20 mK as reference) of the high mode $\Delta P_+$ against period shifts for the low mode $\Delta P_-$. Circles around the data points indicate period values from 50 mK to 600 mK at 50 mK intervals. The upper and lower lines indicate the predictions for the data based on the supersolid and shear stiffening hypotheses.

**4. Analysis of Triple Oscillator:**
The equations of motion for our triple oscillator are:
$$\begin{pmatrix} I_1\omega^2 - k_1 & k_1 & 0 \\ k_1 & I_2\omega^2 - k_1 - k_2 & k_2 \\ 0 & k_2 & I_3\omega^2 - k_2 - k_3 \end{pmatrix} \begin{pmatrix} \theta_1 \\ \theta_2 \\ \theta_3 \end{pmatrix} = \begin{pmatrix} 0 \\ 0 \\ 0 \end{pmatrix}$$

Setting the determinant of this 3×3 matrix to zero leads to a cubic equation in $\omega^2$, with three roots corresponding to the three resonances of the triple oscillator. The roots for a cubic equation can be expressed in closed form, but the expressions are complex. A simpler approach is justified in the present case since the moments of inertia of the helium sample and the internal torsion bob are small compared to the moments of inertia of the cell and the dummy oscillator. We shall treat the system as a double oscillator where a periodic back action torque with amplitude $\tau = \chi(\omega)\theta_2$ acts on the cell in addition to the torque from its own torsion rod. This torque is just that which is required for the angular acceleration of the helium sample and the internal torsion bob during the oscillation of the cell. Then in terms of a double oscillator we have for the equations of motion:

$$\begin{pmatrix} I_2\omega^2 - k_2 + \chi(\omega) & k_2 \\ k_2 & I_3\omega^2 - k_2 - k_3 \end{pmatrix} \begin{pmatrix} \theta_2 \\ \theta_3 \end{pmatrix} = \begin{pmatrix} 0 \\ 0 \end{pmatrix}.$$

The equations of motion then lead to a quadratic equation with two solutions, $\omega_-^2$ and $\omega_+^2$:

$$\omega_\pm^2 = \frac{\left(I_2(k_2+k_3)+I_3(k_2-\chi(\omega_\pm))\right)}{2I_2I_3}\left[1\pm\sqrt{1-\frac{4I_2I_3(k_2k_3-(k_2+k_3)\chi(\omega_\pm))}{\left(I_2(k_2+k_3)+I_3(k_2-\chi(\omega_\pm))\right)^2}}\right].$$

The above equation could be solved by treating $\chi(\omega)$ as a small perturbation.

To estimate the back action torque $\chi(\omega)\theta_2$, we first note that for these two modes, the motion of the internal torsion bob is in phase with that of the cell and $\theta_1 = \theta_2 + \Delta\theta$. The torque stems from the difference between $\theta_1$ and $\theta_2$, which can be expressed as $\chi(\omega)\theta_2 = k_1\Delta\theta = (k_0 + k_\mu)\Delta\theta$. It also determines the angular acceleration of the internal torsion bob and the sample, which means $\chi(\omega)\theta_2 = -\omega^2(I_{He} + I_1)(\theta_2 + \Delta\theta) \approx -\omega^2(I_{He} + I_1)\theta_2$, since $\Delta\theta \ll \theta_2$ which is justified by the fact that $\omega_1 \gg \omega_\pm$. Solving for $\chi(\omega)$, we get:

$$\chi(\omega) = -\omega^2(I_{He}+I_1)\left(1+\frac{\omega^2(I_{He}+I_1)}{k_0+k_\mu}\right) = -\omega^2(I_{He}+I_1)\left(1+\frac{\omega^2(I_{He}+I_1)}{k_0+G\mu}\right).$$

The sensitivity of the periods to variation in mass-loading due to changes in the moment of inertia of the $^4$He solid is:

$$\frac{dP}{dI_{He}} = \frac{dP}{d\chi(\omega)}\frac{d\chi(\omega)}{dI_{He}} \approx -\omega^2\frac{dP}{d\chi(\omega)} = \frac{\delta P}{\Delta I}.$$

We neglected the term $\frac{\omega^2(I_{He}+I_1)}{k_0+G\mu}$ in this process because it is much smaller than 1 – equal to 0.00351 for the $P_-$ mode and 0.02376 for the $P_+$ mode. The ratio of the sensitivities to mass loading for the + and − modes is:

$(dP_+/dI_{He})/(dP_-/dI_{He}) = (\delta P_+/\delta P_-)_{\Delta I} = 0.234$, given by the previous mass loading calibration.

For the sensitivity to changes in the shear modulus we have, instead:

$$\frac{dP}{d\mu} = \frac{dP}{d\chi(\omega)}\frac{d\chi(\omega)}{d\mu} = -\omega^4\frac{dP}{d\chi(\omega)}\frac{G(I_{He}+I_1)^2}{(k_0+k_\mu)^2} = \omega^2\frac{G(I_{He}+I_1)^2}{(k_0+k_\mu)^2}\frac{\delta P}{\Delta I}.$$

Here the ratio of sensitivities for small changes in the shear modulus is

$(dP_+/d\mu)/(dP_-/d\mu) = (P_-/P_+)^2\left[(\delta P_+/\delta P_-)_{\Delta I}\right] = (\delta P_+/\delta P_-)_{\Delta\mu} = 1.526$.

Thus the sensitivity factor for changes in the shear modulus is larger than the factor for mass-loading by the ratio $(P_-/P_+)^2$, or, $(\delta P_+/\delta P_-)_{\Delta\mu} = 1.526$ as compared to $(\delta P_+/\delta P_-)_{\Delta I} = 0.234$ from the mass-loading calibration.

The above analysis also enables us to estimate the change in $\mu$ as the sample is cooled from 500mK to 20mK. The result is a 43.9 percent of increase for the low mode and a 53.6 percent of increase for the high mode.

Figure 3 shows that the actual data agree much more closely with the shear stiffening model, indicating that the dominant cause for the oscillation frequency shifts is changes in shear modulus of $^4$He. Although the small deviation from the predicted slope may indicate that a small fraction of frequency shifts can still be attributed to superfluid-like mass decoupling, the small temperature dependence of the empty cell periods and other background factors may also account for the differences in the slopes.

## 5. Conclusions and Acknowledgements:

We have shown that the NCRI-like period shifts seen with our triple oscillator are incompatible with the classical predictions based on supersolidity, whereas they can be adequately explained as arising from the shear modulus anomaly [4]. In light of these findings, the period shift data of the earlier experiment [5] should be interpreted as a consequence of the temperature dependence of the shear modulus. The substantial increase of the oscillator period in the high temperature region (above 200 mK), following deformation, would then imply a major softening of the high temperature shear modulus, while the modulus returns to the same fixed value at the lowest temperature below 20 mK. The observation of a similar behavior in the anomalous softening of $^4$He crystals has been reported by the Paris group of Balibar and Maris [9]. The next step in our research program will be to use the compound oscillator technique to examine the response of an oscillator designed to be insensitive to shear modulus effects in an attempt to distinguish between a true supersolid signal with the correct frequency dependence and the period shift arising from the shear modulus anomaly [4].

This work has been supported by the National Science Foundation through Grants DMR-060586 and PHY-0758104 and the CCMR Grant No. DMR-0520404.